# VASARI-auto: equitable, efficient, and economical featurisation of glioma MRI

James K. Ruffle[1,2], Samia Mohinta[1], Kelly Pegoretti Baruteau[2], Rebekah Rajiah[1], Faith Lee[1], Sebastian Brandner[3], Parashkev Nachev[1] and Harpreet Hyare[1,2]

[1]Queen Square Institute of Neurology, University College London, London, UK

[2]Lysholm Department of Neuroradiology, National Hospital for Neurology and Neurosurgery, London, UK

[3]Division of Neuropathology and Department of Neurodegenerative Disease, Queen Square Institute of Neurology, University College London, London, UK

Correspondence to:

Dr James K Ruffle

Email: j.ruffle@ucl.ac.uk

Address: Institute of Neurology, UCL, London WC1N 3BG, UK

## Manuscript Type

Original article

## Conflict of interest

None to declare.

## Funding

JKR was supported by the Medical Research Council (MR/X00046X/1), the British Society of Neuroradiology, and the National Brain Appeal. PN is supported by the Wellcome Trust (213038/Z/18/Z), and the UCLH NIHR Biomedical Research Centre. HH is supported by the National Brain Appeal and the UCLH NIHR Biomedical Research Centre.

## Authorship

JKR: conceptualisation, methodology, software, validation, formal analysis, resources, writing – original draft, writing - review & editing, visualisation.

SM: methodology, formal analysis, writing - review & editing.

KPG: validation, formal analysis, writing - review & editing.

RR: methodology, writing – original draft, writing - review & editing.

FL: methodology, writing - review & editing.

SB: validation, resources, writing - review & editing.

PN: methodology, resources, writing – original draft, writing - review & editing.

HH: conceptualisation, methodology, validation, formal analysis, resources, writing – original draft, writing - review & editing.

All authors have been involved in the writing of the manuscript and have read and approved the final version.

## Acknowledgements

We are most grateful for the support of the British Neuro-Oncology Society, which assisted us in accumulating a list of all UK neuro-oncology centres.

## Code, model, and data availability

The software for VASARI-auto shall be openly available upon publication at https://github.com/jamesruffle/vasari-auto. Our tumour segmentation model and code are openly and freely available at https://github.com/high-dimensional/tumour-seg. All patient data utilised in this article is freely and openly available.

## Abstract

The VASARI MRI feature set is a quantitative system designed to standardise glioma imaging descriptions. Though effective, deriving VASARI is time-consuming and seldom used clinically. We sought to resolve this problem with software automation and machine learning. Using glioma data from 1172 patients, we developed VASARI-auto, an automated labelling software applied to open-source lesion masks and an openly available tumour segmentation model. Consultant neuroradiologists independently quantified VASARI features in 100 held-out glioblastoma cases. We quantified 1) agreement across neuroradiologists and VASARI-auto, 2) software equity, 3) an economic workforce analysis, and 4) fidelity in predicting survival. Tumour segmentation was compatible with the current state of the art and equally performant regardless of age or sex. A modest inter-rater variability between in-house neuroradiologists was comparable to between neuroradiologists and VASARI-auto, with far higher agreement between VASARI-auto methods. The time for neuroradiologists to derive VASARI was substantially higher than VASARI-auto (mean time per case 317 vs. 3 seconds). A UK hospital workforce analysis forecast that three years of VASARI featurisation would demand 29,777 consultant neuroradiologist workforce hours and >£1.5 ($1.9) million, reducible to 332 hours of computing time (and £146 of power) with VASARI-auto. The best-performing survival model utilised VASARI-auto features instead of those derived by neuroradiologists. VASARI-auto is a highly efficient and equitable automated labelling system, a favourable economic profile if used as a decision support tool, and non-inferior survival prediction. Future work should iterate upon and integrate such tools to enhance patient care.

## Introduction

Contemporary brain tumour care relies upon joint multi-disciplinary teams spanning clinical, oncological, surgical assessment, histopathology, and radiology(Louis et al., 2021). Neuroradiology plays a vital role for these patients, not merely in the initial diagnosis and triaging to services but also in post-treatment follow-up, where many patients are monitored for several years. Across all subspecialties, our understanding of neuro-oncology is increasingly recognised to be challenged by the marked heterogeneity of brain tumours(Ruffle et al., 2023b). Though there is no established solution to this heterogeneity, it is a problem that could arguably only be attended to with richer patient-personalised information, catalytic for data-driven decision-making. But to understand this heterogeneity, we require robust systems that illuminate disease variation from one patient to another.

The VASARI (Visually AcceSAble Rembrandt Images) MRI feature set is a quantitative scoring system designed to facilitate accurate and reliable imaging descriptions of adult gliomas(TCIA, 2020), initially developed in 2010 as part of The Cancer Genome Atlas (TCGA) initiative from the Repository for Molecular BRAin Neoplasia DaTA (REMBRANDT) study(Gusev et al., 2018). It uses controlled and predefined terminology to define hallmark characteristics of glioma – including location, proportions of constituent components (such as oedema, enhancing and nonenhancing tumour), and other associated features such as cortical, ependymal, or deep white matter involvement.

VASARI's inception was intended to yield more consistent imaging interpretations, irrespective of its rater, centre or imaging approach(TCIA, 2020). Indeed, it has shown promise towards better standardisation of care in adult glioblastoma, with multiple studies consistently demonstrating reasonable inter-observer agreement across constituent VASARI features beyond what could be expected from a conventional means of reporting(Gemini et al., 2023; Park et al., 2021; Setyawan et al., 2024). It has also been used with clinical and genomic data to effectively predict tumour histological grade, progression, mutation status, risk of recurrence and overall patient survival, implying a broader clinical utility(Jain et al., 2014; Nicolasjilwan et al., 2015; Peeken et al., 2019; Peeken et al., 2018; Setyawan et al., 2024; Wang et al., 2021; Zhou et al., 2017). Though initially developed for adult glioblastoma, the VASARI feature set has been trialled in several novel clinical contexts, including in paediatric brain tumours(Biswas et al., 2022) and rarer neuroepithelial malignancies(Li et al., 2023), where it has shown potential as a clinical aid.

However, despite good evidence to support implementing the VASARI feature set as a clinical tool, it can be prohibitively time-consuming. Some studies report manual segmentation times of 20-40 minutes per case(Deeley et al., 2011; Wan et al., 2020). In an inevitably resource-limited and overstretched healthcare system(NHS, 2024), such a time constraint inevitably obstructs translation into real-world care.

Though the task is complex, it is theoretically deliverable by machine vision. Over the last few decades, lesion segmentation has formed a cornerstone of innovation across neuro-oncology(Lu et al., 2021; Peng et al., 2021; Ruffle et al., 2023a; Xue et al., 2020), medical imaging(Lenchik et al., 2019; Suetens et al., 1993), biomedical engineering(Ashburner and Friston, 2005), machine and deep learning(Menze et al., 2015). The ability to segment an anatomical or pathological lesion in three dimensions confers the ability to evaluate it quantitatively – moving beyond visual qualitative assessment – with greater richness and fidelity than conventional two-dimensional measurements repeatedly shown to be often spurious and inconsistent between radiologists(Dempsey et al., 2005; McNitt-Gray et al., 2015; Zhao et al., 2009), and with greater sensitivity to the heterogeneity of the underlying pathological patterns(Mandal et al., 2020). Enabling radiological image segmentation opens many possibilities for downstream innovation in neuro-oncological healthcare and research, ranging from standardisation of care, clinical stratification, outcome prediction, response assessment, treatment allocation and risk quantification, many of which have already shown great promise. The underlying goal is to enhance the individual fidelity of data-driven decision-making, facilitating better patient-centred care(Rajpurkar et al., 2022; Topol, 2019), a remit especially warranted in neuro-oncology(Louis et al., 2021).

Given this, we developed VASARI-auto, an automated VASARI feature set labelling tool (Figure 1). With a required input of patient lesion segmentations only – engineered by design to maximise patient confidentiality - we herein illustrate its high performance, efficiency, equity, and downstream survival predictive utility in a multi-site patient cohort large-in-kind, and with real-world healthcare provider simulations illustrating tangible added value that can enhance clinical neuro-oncology workflows.

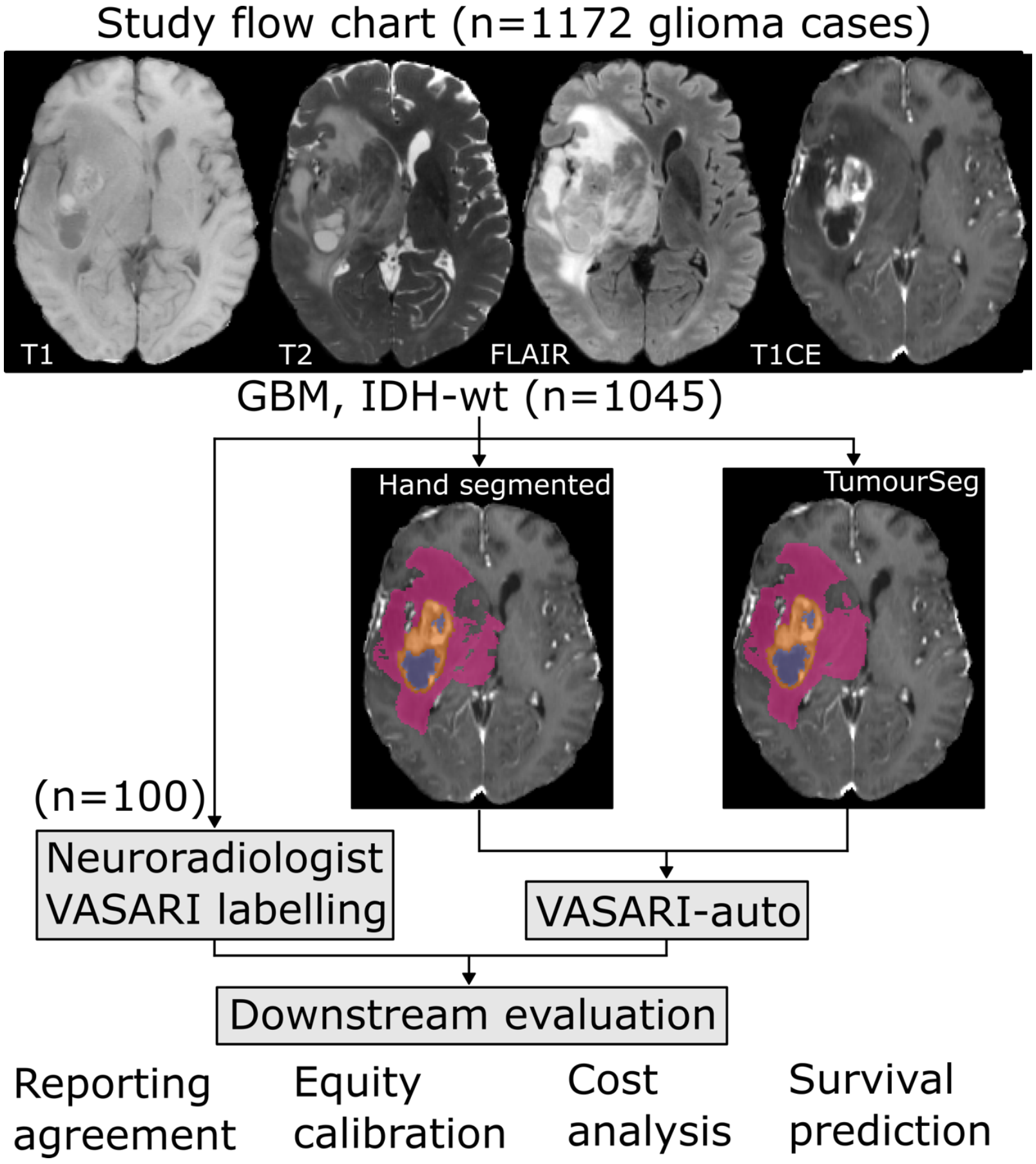

**Figure 1. Study flow chart.** Volumetric T1, T2, FLAIR, and post-contrast T1-weighted imaging were acquired for all participants. By random assignment of 100 glioblastoma, IDH-wt cases, two experienced consultant neuroradiologists reviewed imaging and recorded VASARI features and were timed doing so. In parallel, we developed VASARI-auto, an automated software to determine VASARI features. We derived these features using VASARI-auto from both semi-supervised hand-annotated lesion masks from a separate group of neuroradiologists and using a previously published and openly available tumour segmentation model, herein referred to as 'TumourSeg'. Lesional tissue is colour-coded as orange for enhancing tumour, purple for nonenhancing tumour, and pink for perilesional signal change. We subsequently undertook multiple downstream evaluations of both neuroradiologist VASARI labelling and that from VASARI-auto, evaluating: 1) agreement both between neuroradiologists, between software, and between neuroradiologist and software; 2) equity calibration to determine if neuroradiologist and software labelling were equitably performant for all ages and sexes; 3) a simulated economic analysis determining the cost to undertake labelling with neuroradiologists or VASARI-auto based on real-world clinical workloads; and 4) in using these data to predict patient overall survival. Neuroradiologists were blinded to all software development and evaluations from VASARI-auto, and likewise, software developers were blinded to all neuroradiologist labelling until the final downstream evaluation stage.

## Materials and Methods

### Data

We utilised open-source neuro-oncology pre-treatment patient imaging data (n=1172) provided by The University of California San Francisco Preoperative Diffuse Glioma MRI (UCSF-PDGM) (n=501)(Calabrese et al., 2022) and the University of Pennsylvania Glioblastoma Imaging, Genomics, and Radiomics (UPenn-GBM) (n=671)(Bakas et al., 2022b) repositories. Data were acquired across multiple scanners and acquisition protocols, with further details provided by the curators online(Bakas et al., 2022a; Calabrese et al., 2022).

We firstly contacted the corresponding authors of both datasets to clarify which participant imaging was part of The Brain Tumour Segmentation Challenge (BraTS)(Baid et al., 2021) since BraTs data were used to initially train the adopted tumour segmentation model (herein referred to as 'TumourSeg')(Ruffle et al., 2023a; Ruffle, 2023), and as such needed to prevent any possibility of an information leak. We excluded any such cases from the data pool. Further, we subsampled to study only patients with a confirmed molecular diagnosis of glioblastoma, IDH-wt(Louis et al., 2021), for which VASARI featurisation was initially developed. Each patient dataset included volumetric and brain-extracted T1, T2, FLAIR, and post-contrast T1-weighted MRI sequences (Table 1).

In all patients, age, sex, overall survival (in days), and a lesion segmentation were available. Separately undertaken by the original UCSF-PDGM and UPenn-GBM repository authors, each patient neuroimaging set first underwent automated segmentation using an ensemble model consisting of the prior top-scoring BraTS challenge algorithms, which was then manually corrected by a group of annotators with varying experience and approved by one of two neuroradiologists with more than 15 years of attending experience each(Bakas et al., 2022b; Calabrese et al., 2022).

### Ethical approval

UCSF-PDGM data collection followed relevant guidelines and regulations and was approved by the UCSF institutional review board with a waiver for consent(Bakas et al., 2022b). For UPenn-GBM, collection, analysis, and release of the UPenn-GBM data was approved by the Institutional Review Board at the University of Pennsylvania Health System (UPHS), and informed consent was obtained from all participants(Bakas et al., 2022b).

## Neuroimaging

### Neuroradiologist VASARI-featurisation

From this glioblastoma, IDH-wt cohort, we drew a random sample of 100 unique patients. The delineation of 100 from 1172 patients for manual neuroradiologist labelling was programmatically randomised to avoid selection bias. Our choice of n=100 was guided by a balance of reasonable statistical power and how time-consuming manual annotation of these scans can be for neuroradiologists.

Each patient was randomly assigned to one of two consultant neuroradiologists with 15 and 8 years of experience in neuro-oncology, who quantified VASARI features from neuroimaging. Both radiologists had prior experience with VASARI criteria, though an initial calibration meeting was also undertaken to ensure consistency. Neuroradiologists quantified all VASARI features except those requiring diffusion or non-brain-extracted sequences. The time taken to derive VASARI features in each patient case was recorded. From a random number generator, we drew a random integer between 10-15 (which drew 13), for which we randomly allocated 13 duplicate cases to both neuroradiologists to ascertain inter-rater agreement. Neuroradiologists were blinded to all software and model development.

### Tumour segmentation

We used a previously published tumour segmentation model (TumourSeg) for all cases, described in significant detail elsewhere(Ruffle et al., 2023a). In brief, this model is a high-resolution convolutional 3D U-Net implemented with nnU-Net(Isensee et al., 2021), a pipeline with proven high performance in semantic segmentation across a range of micro and macroscopic tasks(Antonelli et al., 2022; Isensee et al., 2020; Isensee et al., 2021). The model was trained on the BraTS training dataset of 1251 participants with 5-fold cross-validation, with additional external evaluation on cases acquired at the National Hospital for Neurology and Neurosurgery(Ruffle et al., 2023a). We ensured that none of the BraTS data used in model training were evaluated in this downstream task to prevent the possibility of an information leak. We compared the segmentation performance of TumourSeg to hand-annotated labels provided by the original dataset curators: quantitatively by the Dice-Sørensen coefficient and qualitatively by a neuroradiologist's visual review.

### Nonlinear registration with enantiomorphic normalisation

Having segmented lesions in native space, MRI sequences and lesion segmentation masks were nonlinearly registered to 1mm MNI space with Statistical Parametric Mapping (SPM) using enantiomorphic correction(Ashburner and Friston, 1999; Nachev et al., 2008). The advantage of enantiomorphic correction is that the risks of registration errors secondary to a lesion are minimised by leveraging a given patient's normal structural neuroanatomy on the unaffected contralesional hemisphere(Nachev et al., 2008). A neuroradiologist manually reviewed all imaging data at multiple stages of the data pre-processing.

## VASARI-auto software development

We developed a fully automated pipeline – 'VASARI-auto' – to derive VASARI features from lesion masks. Lesion masks could be of any source, whether manually traced, from an openly available tumour segmentation model, or other lesion segmentation tools. VASARI-auto required data to be held in MNI registered space (prototyped in a 1mm$^3$ volumetric resolution, but deployable in any). We pooled neuroanatomical atlases for all brain lobes, as well as the brainstem, insula, thalamus, corpus callosum, internal capsule, ventricles, and cortex, for the derivation of locational-based features. For each case, VASARI-auto loaded the multi-channel tumour segmentation (with separate labels for enhancing tumour, nonenhancing tumour, and perilesional signal change) and, following pre-existing VASARI reporting standards(TCIA, 2020), derived the following: F1 - tumour location; F2 – side of tumour epicentre; F4 – enhancement quality; F5 – proportion enhancing; F6 – the proportion of nonenhancing tumour; F7 – the proportion of necrosis; F9 – multifocal/multicentric lesional status; F11 – the thickness of the enhancing margin; F14 – the proportion of oedema; F19 – ependymal invasion; F20 – cortical involvement; F21 – deep white matter invasion; F22 whether nonenhancing tumour crossed the midline; F23 – whether enhancing tumour crossed the midline; and F24 – the presence of satellite lesions. Notably, whilst initial tumour segmentation harnesses a trained, validated, and open-sourced deep learning segmentation model, VASARI-auto requires only mathematical derivation of features from the 3-dimensional lesion mask. No non-deterministic or nonlinear inferential statistics are involved, and the results are mathematically deterministic and reproducible.

Our code did not quantify a few VASARI features that require either non-brain extracted data (F25 – calvarial modelling) or the original MRI sequences (F10 T1/FLAIR ratio; F12-13 – definition of the enhancing and nonenhancing margin, F18 – pial invasion, and F16 – haemorrhage), the reason being was that we wished to develop an automated tool immediately usable with irrevocably anonymised lesion segmentation data without the requirement for raw volumetric neuroimaging. We similarly did not quantify F17 - diffusion changes since DWI was not available for many cases in the external data, beyond our control. We also did not quantify the presence of F8 – cysts since most brain tumour segmentation models rely on BraTS lesion labels of enhancing tumour, nonenhancing tumour, and perilesional signal change, but with no distinction for cysts. Therefore, we felt any attempts to model cyst presence would be liable to confabulation. We similarly did not quantify F3 – eloquence, for lack of appropriate brain masks to model it robustly; moreover, we did not wish to detract from a gold standard of a neurosurgeon's electrical stimulation assessment for eloquent-sparing resections(Ritaccio et al., 2018).

The requirements to run VASARI-auto are given below in the software subsection. We also recorded time to quantify VASARI features with VASARI-auto, both already pre-generated lesion masks, and when paired with TumourSeg(Ruffle et al., 2023a).

## Downstream evaluation

### Reporting agreement

Quantitatively, we compared agreement in all VASARI featurisation between 1) consultant neuroradiologists, 2) consultant neuroradiologists and VASARI-auto, and 3) between VASARI-auto when using either the source semi-supervised and neuroradiology-reviewed segmentations to VASARI-auto using TumourSeg(Ruffle et al., 2023a; Ruffle, 2023). The neuroradiologist's label was always taken as the ground truth against which VASARI-auto would be assessed. The agreement was quantified by Cohen's Kappa(Pedregosa et al., 2011), which furthermore was appropriately linearly weighted for non-Boolean VASARI features. We also quantified the balanced accuracy in VASARI featurisation between consultant neuroradiologists (the ground truth) and VASARI-auto (the prediction), as well as the balanced accuracy between VASARI-auto using the source semi-supervised and neuroradiology-reviewed segmentations (the ground truth) and VASARI-auto using TumourSeg (the prediction)(Ruffle et al., 2023a).

Qualitatively, in post hoc analyses, we also undertook a case-based review of 1) the results from TumourSeg, with direct comparison to the neuroradiologist hand annotation, and 2) the results of VASARI-auto with direct comparison to the VASARI featurisation of separate neuroradiologists.

### Equity calibration

We quantified software and reporting patient equity(Abramoff et al., 2023; Carruthers et al., 2022) for all analysis steps. For tumour segmentation, we compared model performance by the Dice coefficient across all lesional compartments (enhancing tumour, nonenhancing tumour, perilesional signal change, and whole tumour [a single mask for all areas of abnormality]) for male and female sex and for all decades of age included in the cohort (20-90 years). We similarly compared Cohen's Kappa agreement metrics across male and female sex and all decades of age.

### Efficiency, economic and workforce analysis

We statistically compared the time required to record VASARI features between 1) consultant neuroradiologists, 2) VASARI-auto with tumour segmentations already supplied, and 3) VASARI-auto with TumourSeg.

Next, we undertook an economic and workforce analysis, simulating neuro-oncology workload across the UK. Every week, a neuro-oncology multidisciplinary team (MDT) meeting is held to discuss all referrals, ongoing cases, and management plans, which includes a neuroradiological review of all cases. We reviewed the last three years of neuro-oncology MDT lists (2020-2023) and quantified the minimum and maximum number of cases to be discussed each week, which was of range 30-75. We determined the minimum and maximum pay scales for consultants in the National Health Service workforce as of March 2024, which vary depending on years of service(NHS-Employers, 2023). We similarly quantified power consumption costs to run a reasonably powerful computer (1200 kilowatt Hour (kWh), based on UK energy tariffs as of March 2024(sust-it.net, 2024). We curated a list of all UK neuro-oncology centres (n=40), kindly provided by the British Society of Neuro-Oncology, to simulate UK-wide neuro-oncology workloads.

Having derived this data, we simulated the next three years (2024-2027) of MDT clinical workload at each centre. A random number of MDT cases was simulated weekly using the previous minimum-maximum caseload through 2020-2023. We then simulated a random choice of neuroradiology consultants who would be allocated to present a given week's neuro-oncology MDT, with their salary drawn randomly from the NHS consultant pay scales. We then randomly simulated the time taken to quantify VASARI features across all cases, where time per case was drawn from a random uniform distribution informed by the time taken for neuroradiologists to quantify all 100 cases in our earlier analysis. From this, we quantified the workload and financial cost if each patient had undergone VASARI featurisation by a neuroradiologist. We similarly quantified the time and expense of power if VASARI-auto and VASARI-auto with TumourSeg had undertaken featurisation. We undertook this process with five iterations to ensure model stability/robustness to outliers.

### Survival prediction

Lastly, we fitted linear regression models seeking to predict patient overall survival (OS) (in days) from VASARI features. These were in the formulation $OS \sim 1 + f_1 + f_2 + f_n$, where $f_n$ denotes each VASARI feature. We fitted separate models using VASARI features quantified from 1) consultant neuroradiologists, 2) VASARI-auto using the source semi-supervised and neuroradiology-reviewed segmentations, and 3) VASARI-auto using TumourSeg(Ruffle et al., 2023a; Ruffle, 2023), from which we compared the quality of fit. We derived each feature's variance inflation factor to adjust for potential multicollinearity and excluded those whose value exceeded 10. Although large in kind (n=100), with the relatively small sample used here, we deliberately chose not to model with nonlinear or machine learning models nor partition data into train or test datasets, which would otherwise be highly liable to overfit in such an instance. Since our task here is to benchmark the utility of features derived by neuroradiologists compared to our developed machinery, using nonlinear complex models that are liable to overfit would arguably be inappropriate in this setting.

## Analytic compliance

All analyses were performed and reported following international TRIPOD and PROBAST-AI guidelines(Collins et al., 2021).

## Code, model, and data availability

The software for VASARI-auto shall be openly available upon publication at https://github.com/jamesruffle/vasari-auto. All patient data utilised in this article is freely and openly available(Bakas et al., 2022b; Calabrese et al., 2022).

## Software

The following software and models were used for analyses: Matplotlib(Hunter, 2007), MONAI(The-MONAI-Consortium, 2020), Nibabel(Brett et al., 2020), Nilearn(Nilearn-contributors, 2024), NumPy(Harris et al., 2020), pandas(Reback et al., 2020), PyTorch(Paszke et al., 2019), seaborn(Waskom and Seaborn-Development-Team, 2020), scikit-learn(Pedregosa et al., 2011), and TumourSeg(Ruffle et al., 2023a; Ruffle, 2023), all within a Python environment.

## Compute

All experiments were performed on a 64-core Linux workstation with 256Gb of RAM and an NVIDIA 3090Ti GPU.

# Results

## Cohort

The brain tumour patient cohort included 56 male and 44 female participants, with a mean age ± standard deviation of 61 years ± 13.49. The mean overall survival (in days) was 436 ± 462.21 days. Seventy-four participants were included from UPenn-GBM, and 26 were included in UCSF-PDGM. There were no significant differences in age, sex, or survival between participants at either site, indicating a well-standardised and representative multi-site sample.

## Segmentation

A comparison of tumour segmentation TumourSeg(Ruffle et al., 2023a; Ruffle, 2023) to the externally curated semi-supervised labels showed a mean Dice segmentation performance of 0.95 ± 0.05 for the whole tumour, 0.89 ± 0.07 for the enhancing tumour, 0.86 ± 0.11 for the nonenhancing tumour, and 0.91 ± 0.06 for the perilesional signal change. A visual overlay of lesion segmentations to the brain showed no spatial discrepancy (Figure 2). There was no significant difference in segmentation performance between the male and female sexes and across all decades of age, indicating an equitable tumour segmentation model (Figure 3).

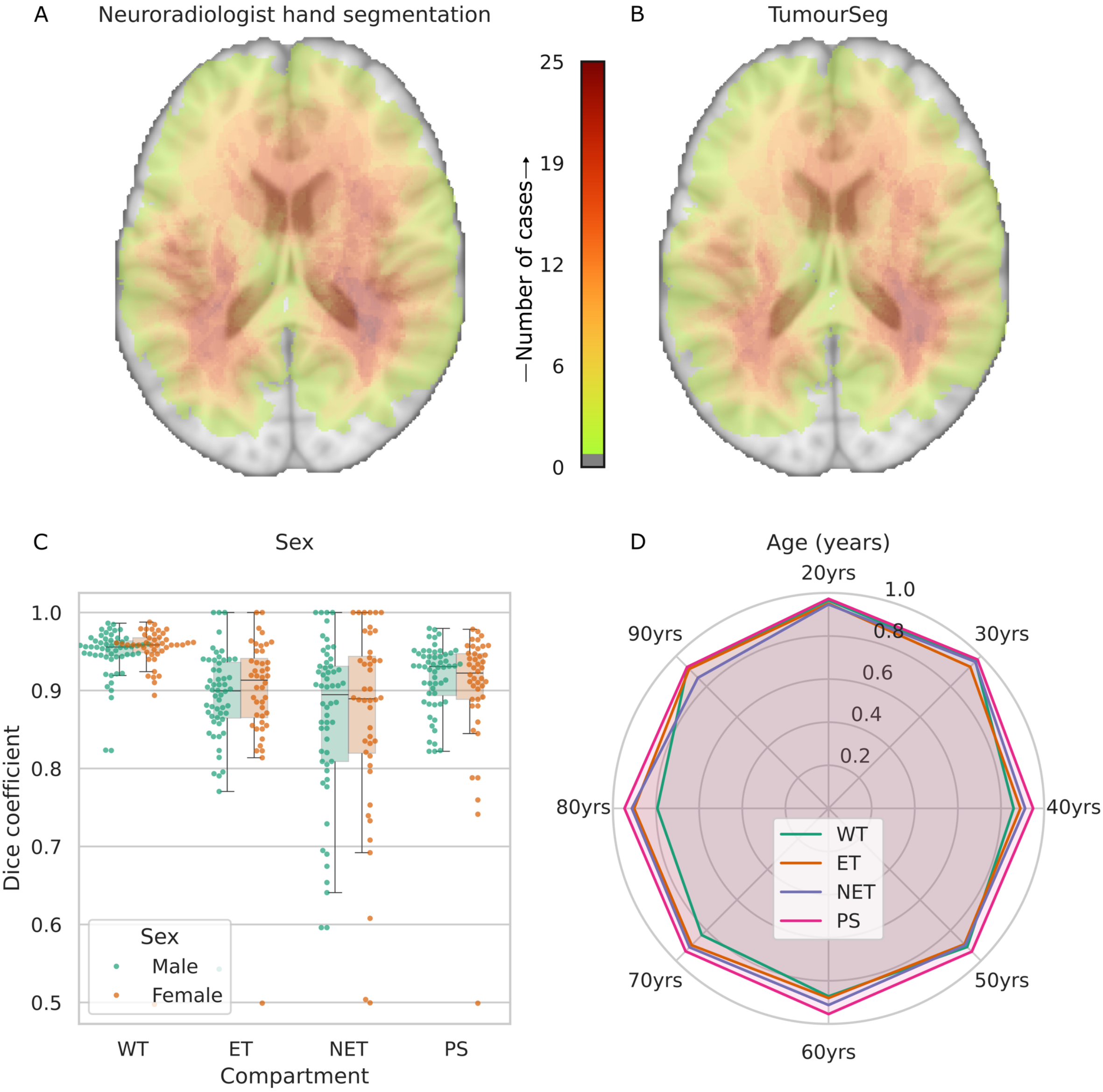

**Figure 2. Tumour segmentation equitable calibration.** A-B) Heatmap of tumour location derived from semi-supervised external neuroradiologist hand segmentations (A) and from segmentation model TumourSeg (B) shows the two to be highly similar indicative of spatially equitable intracranial performance. C-D) Box and whisker (C) and radar (D) plots depict tumour segmentation model performance by Dice coefficient across whole tumour (WT), enhancing tumour (ET), nonenhancing tumour (NET), and perilesional signal change (PS), illustrating that tumour segmentation is equally performant across both male and female patients (C), and across all decades of life (D).

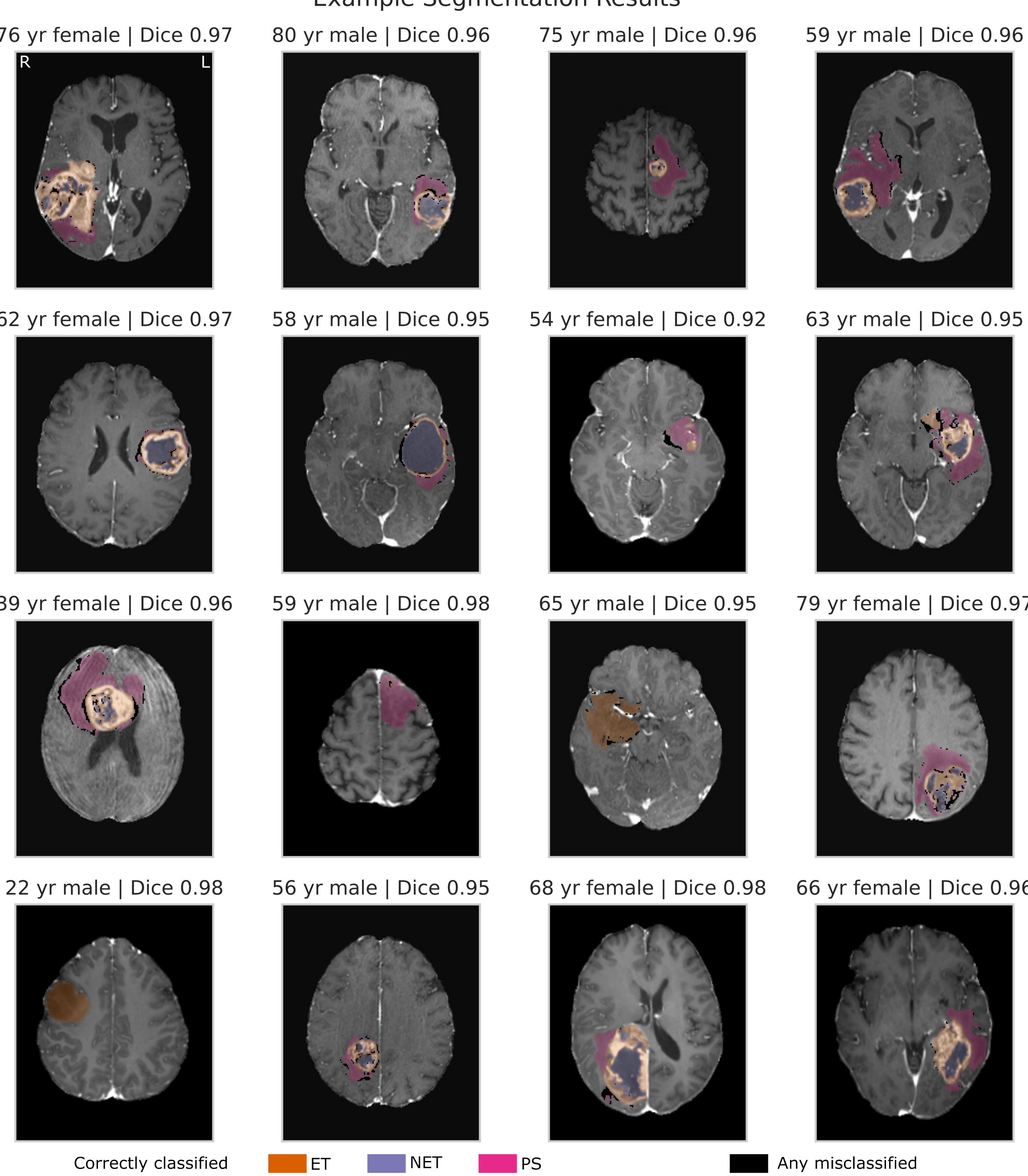

**Figure 3. Randomised case-based review of tumour segmentation**. Randomly selected sample of 16 patients of different ages and sex, with their contrast-enhanced T1-weighted imaging and the TumourSeg model result overlaid. Correctly segmented lesional voxels are colour-coded as orange for enhancing tumour (ET), purple for nonenhancing tumour (NET), and pink for perilesional signal change (PS). Any misclassified voxels (whether false positive, false negative, or correctly lesional but the wrong tissue class) are colour-coded in black.

## Agreement and accuracy evaluations

There was a modest inter-rater agreement between consultant neuroradiologists, with a mean Cohen's Kappa of 0.49 ± 0.32 (Figure 4). The features of highest agreement between neuroradiologists were: F9 – whether a lesion was multifocal/multicentric (Cohen's Kappa 1.00); F20 – cortical involvement (Cohen's Kappa 0.71); and F21 – deep white matter invasion (Cohen's Kappa 0.71). The features of least agreement between neuroradiologists were F24 – presence of satellite lesions (Cohen's Kappa -0.23), F11 – thickness of an enhancing margin (Cohen's Kappa -0.03); and F19 – ependymal invasion (Cohen's Kappa 0.33). Agreement between neuroradiologists and VASARI-auto was relatively similar, with a mean Cohen's Kappa of 0.42 ± 0.34. The features of highest agreement between neuroradiologists and VASARI-auto were: F9 – multifocal/multicentric (Cohen's Kappa 1.00); F2 – side of the epicentre (Cohen's Kappa 0.93), and F1- tumour location (Cohen's Kappa 0.75). The features of least agreement between neuroradiologists and VASARI-auto were F6 – the proportion of nonenhancing tumour (Cohen's Kappa -0.16), F14 – the proportion of oedema (Cohen's Kappa 0.02), and F20 – cortical involvement (Cohen's Kappa 0.11). Agreement between VASARI-auto featurisations, whether using externally curated lesion segmentations or TumourSeg, was substantially higher, with a mean Cohen's Kappa of 0.94 ± 0.10. For this comparison, all feature agreement was 0.88 or higher, apart from F24 – the presence of satellite lesions (Cohen's Kappa 0.59).

Treating neuroradiologist labels as the ground truth, VASARI-auto achieved a mean accuracy of 66% ± 21% (Figure 4). The greatest accuracy was in F2 – side of tumour epicentre (96.55%), F24 – the presence of satellite lesions (86.20%), and F1 – tumour location (80.46%). The lowest accuracy with VASARI-auto was in F6 – the proportion of nonenhancing tumour (20.69%), F14 – the proportion of oedema (32.18%), and F5 – the proportion of enhancing tumour (49.42%). In contrast, when treating VASARI-auto when derived from the external semi-supervised lesion labels as a ground truth, VASARI-auto accuracy using the tumour segmentation model was much more stable, with a mean accuracy of 97.40 ± 0.03%. Case-based examples are shown in Figure 5.

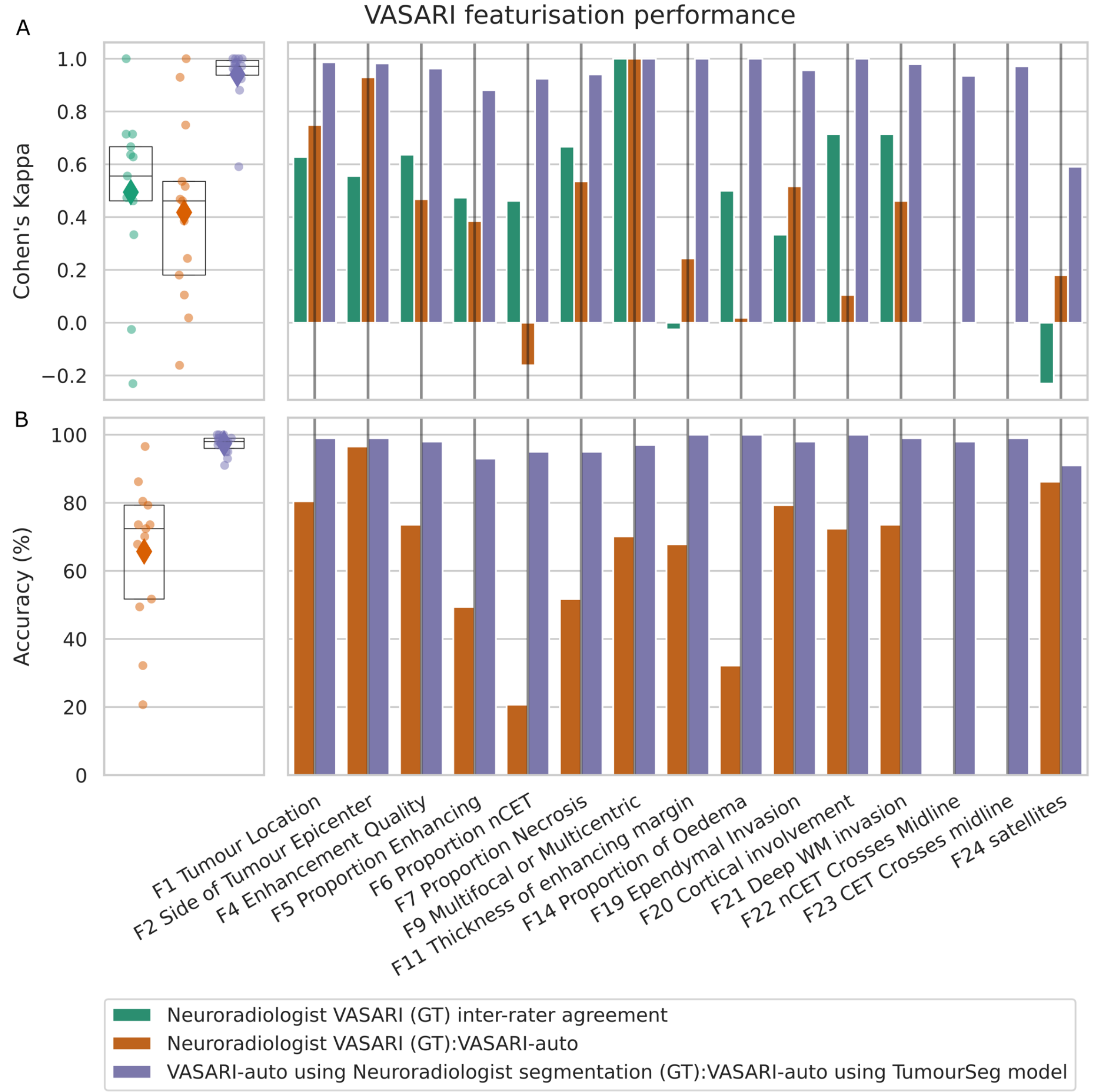

**Figure 4. VASARI featurisation performance.** A) Cohen's Kappa agreement between neuroradiologist reporters (green), between neuroradiologists and VASARI-auto (orange), and between VASARI-auto with and without using TumourSeg (purple) shows inter-rater variability between neuroradiologists but quantitatively higher agreement and stability between both VASARI-auto methods. B) Accuracy between neuroradiologist VASARI reporting (the ground truth, GT) and VASARI-auto (orange), and between VASARI-auto with and without using TumourSeg (purple). VASARI-auto is generally performant compared to neuroradiologists, although some discrepancies are evident due to diverging definitions of what is referred to as a nonenhancing tumour and what is oedema. VASARI-auto across-model comparison is highly accurate. Abbreviations: nCET, non-contrast-enhancing tumour; WM, white matter.

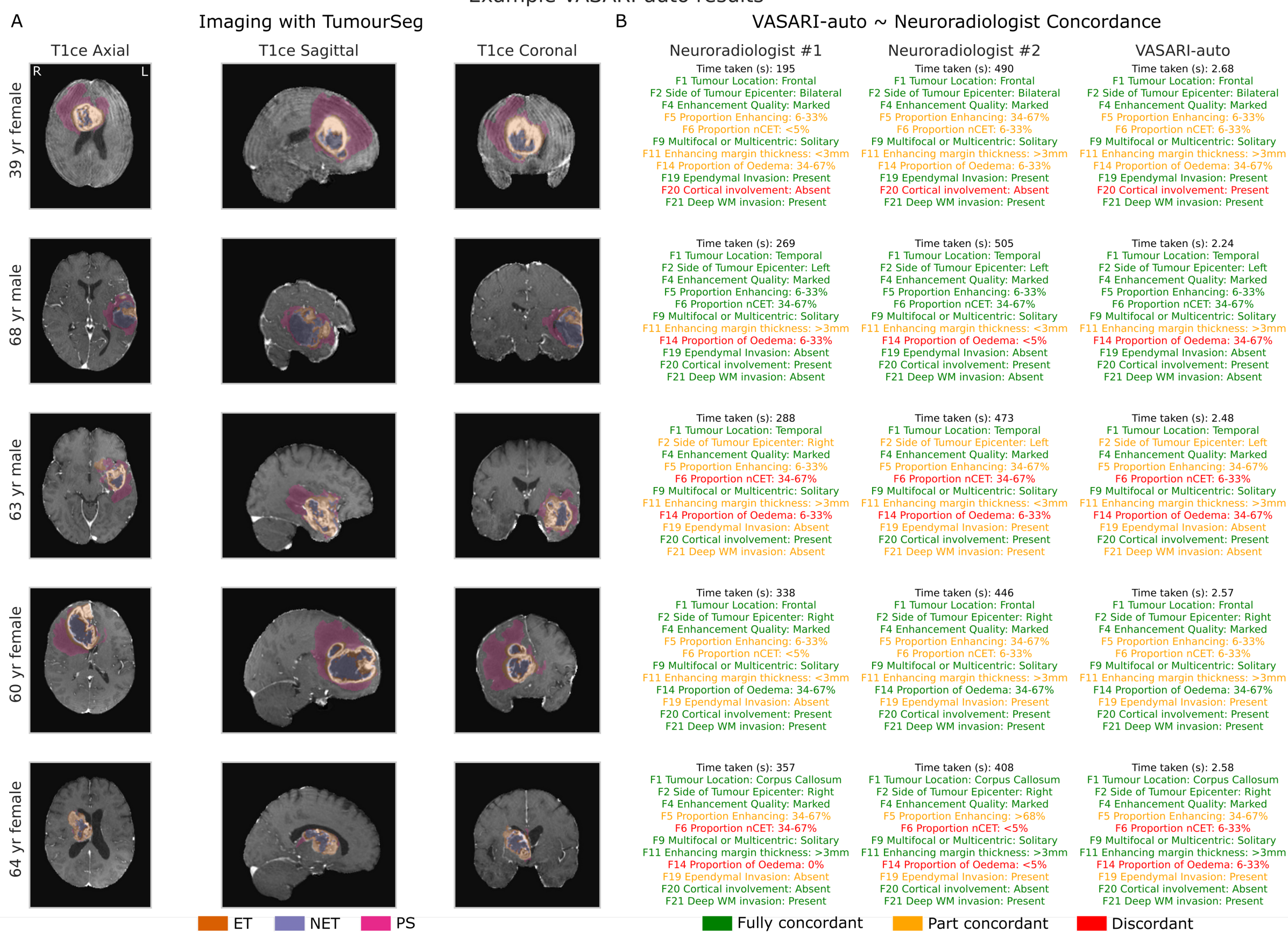

**Figure 5. Randomised case-based review of VASARI featurisation**. A) A randomly selected sample of 5 patients of different ages and sexes, with their contrast-enhanced T1-weighted imaging and the TumourSeg model result overlaid. Correctly segmented lesional voxels are colour-coded as orange for enhancing tumour (ET), purple for nonenhancing tumour (NET), and pink for perilesional signal change (PS). B) Sample VASARI featurisation comparison between Neuroradiologist #1, Neuroradiologist #2, and VASARI-auto. For each case, the time taken to record is listed (in seconds), followed by a selection of VASARI features that are colour-coded depending on whether there is full concordance between both neuroradiologists and VASARI-auto (green), partial concordance between VASARI-auto and one neuroradiologist (orange), or discordant (red).

## Efficiency

The use of VASARI-auto in VASARI featurisation – regardless of whether used in isolation or when paired with the tumour segmentation model – was significantly faster per case compared to consultant neuroradiologists (p<0.0001) (Figure 6). The mean time to quantify was 3.03 ± 0.59 seconds with VASARI-auto, which was significantly higher but notably still efficient at 15.47 ± 1.56 (95%CI) using VASARI-auto with TumourSeg ($p<0.0001$). In comparison, the mean time to quantify was 317.46 (i.e., 5.28 minutes) ± 96.89 seconds with consultant neuroradiologists.

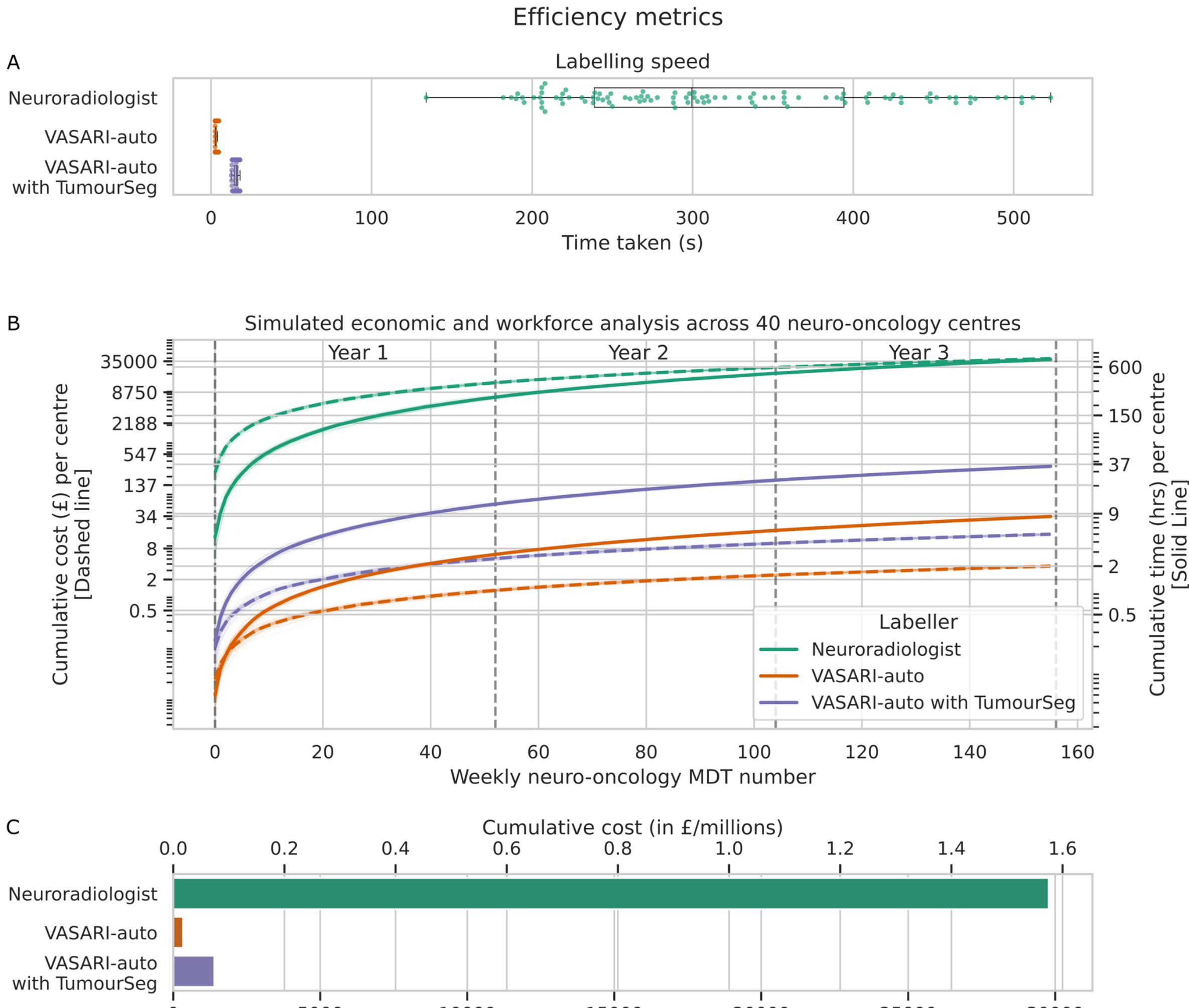

**Figure 6. Efficiency, economic and workforce planning analysis**. A) The time taken for a neuroradiologist to derive a VASARI feature set for a single patient (green) is substantially higher than with either VASARI-auto (orange) or using VASARI-auto paired with TumourSeg (purple). B) Simulated economic and workforce analysis, where weekly neuro-oncology multidisciplinary team (MDT) workload is drawn from a random uniform distribution based upon the last three years of workload at our centre. Thin individual lines represent different simulation runs to emulate the forty different UK neuro-oncology centres, with thicker lines representing the epoch mean. Cumulative financial cost (dashed line) and time taken (solid

line) for a neuroradiologist (green), VASARI-auto (orange), and VASARI-auto with TumourSeg (purple) to derive VASARI features for each patient for each week of the neuro-oncology MDT. The financial cost for consultant neuroradiologists is drawn from a uniform distribution of current National Health Service consultant pay scales provided by the British Medical Association. The power cost for VASARI-auto is based upon current energy prices for a modest 1200 Kilowatt-hour (kWh) GPU-supported computer. Note both y-axes are logged. C) Total UK-wide cumulative cost in time and resource for a neuroradiologist (green), VASARI-auto (orange), and VASARI-auto with TumourSeg (purple) to derive VASARI features for each patient.

## Simulated workforce analysis

The simulated workforce analysis forecast that, over 2024-2027, a total cumulative 8150 ± 168 cases would require discussion at each weekly neuro-oncology MDT (Figure 6). For VASARI featurisation to be undertaken in all cases, this would demand 744.43 ± 15.54 consultant neuroradiologist workforce hours, equating to £39,373.37 ± £864.22 in salary remuneration for hours worked. In contrast, quantifying VASARI features with VASARI-auto for all cases over three years would require 8.30 ± 0.15 hours of computing time (time comparison $p<0.0001$), equating to approximately £3.65 ± 0.12 for power costs (cost comparison $p<0.0001$). If combined with tumour segmentation, this time and expense would rise slightly to 34.89 ± 0.70 hours of computing time and £15.17 ± 0.55 for power costs (both of which remained significantly less than with neuroradiologist labelling). Time taken and costs remained significantly greater for featurisation by neuroradiologists compared to VASARI-auto ($p<0.0001$).

We scaled this up to all 40 neuro-oncology centres across the UK. For VASARI featurisation to be undertaken in all UK cases, this would demand 29,777.39 consultant neuroradiologist workforce hours, equating to £1,574,935 in salary remuneration for hours worked. In contrast, quantifying VASARI features with VASARI-auto for all cases over three years would require 331.95 hours of computing time, equating to approximately £145.85 for power costs. If combined with tumour segmentation, this time and expense would rise slightly to 1394.42 hours of computing time and £606.75 for power costs. Both time taken and cost were significantly greater for featurisation by neuroradiologists compared to VASARI-auto with or without the addition of TumourSeg (all p<0.0001).

## Performance equity

A critical performance measure of any automated tool in healthcare is invariance across patient background characteristics(Carruthers et al., 2022). We compared reporting agreement between 1) neuroradiologists, 2) neuroradiologists and VASARI-auto, and 3) VASARI-auto when applied to the externally curated tumour segmentations or with TumourSeg(Ruffle et al., 2023a), with respect to patient age and sex, using Cohen's Kappa (Figure 7). There was no evidence of reporting inequity between neuroradiologists and between neuroradiologists and VASARI-auto (allowing for the more limited distribution of demographics for those patients double-reported by neuroradiologists). Similarly, agreement between VASARI-auto using manually traced or model-derived lesion segmentations was equally performant across patient age and sex, all of which indicate equitably of VASARI-auto and tumour segmentation models.

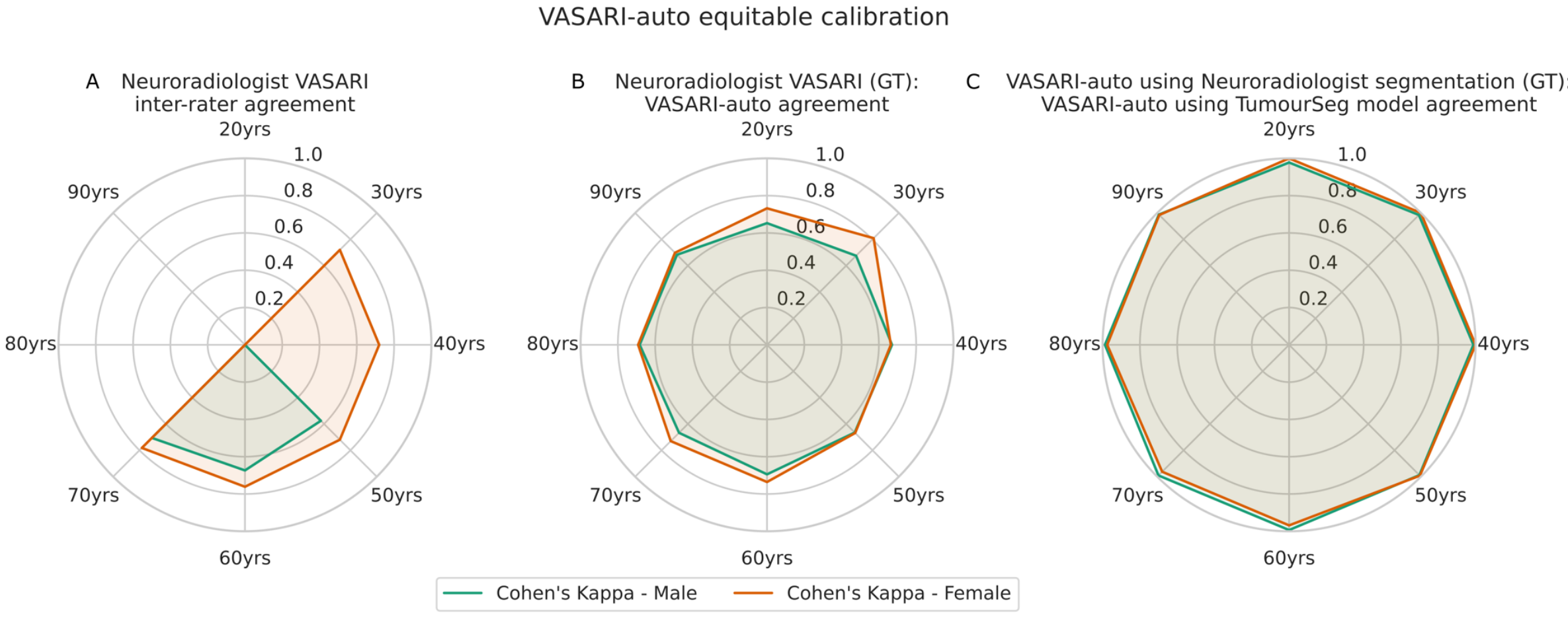

**Figure 7. VASARI featurisation equitable calibration.** A-C) radar plots showing Cohen's Kappa agreement aligned to male (green) or female (orange) patient sex across all decades of life. A) inter-rater agreement between neuroradiologists, B) between neuroradiologists and VASARI-auto, and C) between VASARI-auto using manually traced and software-derived lesion segmentations.

## Survival prediction

The clinical utility of any feature is ultimately determined by its downstream predictive, prescriptive, or inferential power. Fidelity in overall survival prediction using VASARI features was qualitatively similar whether using feature sets derived by consultant neuroradiologists, from VASARI-auto applied to the semi-supervised and neuroradiology reviewed segmentations, or VASARI-auto paired with TumourSeg (Figure 8). Quantitatively, the best linear regression fit was achieved with VASARI-auto using the semi-supervised and neuroradiology-reviewed segmentations ($R^2$ 0.245), followed closely by VASARI-auto using TumourSeg ($R^2$ 0.227), with slightly weaker performance when using the consultant neuroradiologist-labelled VASARI features ($R^2$ 0.205).

Feature-wise, F21 deep white matter invasion was significantly associated with poorer overall survival (p=0.028). Trends for a greater proportion of enhancing tumour, a parietal location, and multifocality were all associated with poorer overall survival, albeit non-significant (p=0.173, p=0.109, and p=0.131, respectively). Full model coefficients are provided in the supplementary material.

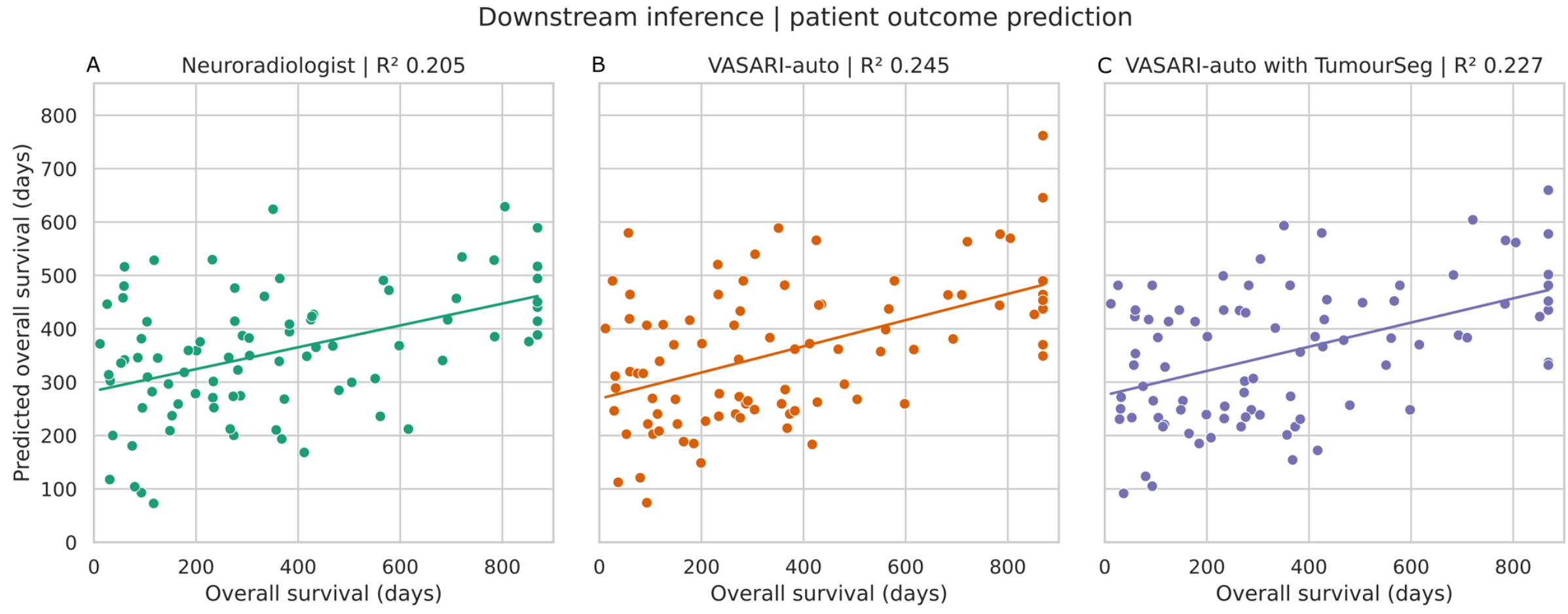

**Figure 8. Downstream inference with patient outcome prediction.** Results of linear regression predicting overall patient survival in days using VASARI-features derived by A) neuroradiologists (green), B) VASARI-auto from the semi-supervised external segmentations (orange), or C) VASARI-auto combined with TumourSeg (purple). X-axes illustrate the actual survival, whereas y-axes illustrate predicted survival. There is highly similar qualitative performance in survival prediction regardless of whether a neuroradiologist or VASARI-auto labels it, although quantitatively, the $R^2$ is higher with both VASARI-auto assessments.

## Discussion

We present VASARI-auto, an automated system for deriving VASARI features from glioma imaging using tumour segmentations alone. Our evaluation shows high accuracy, greater consistency than inter-agreement between neuroradiologists, and equitable performance across age and sex. We show VASARI-auto could save time and resources within each radiology department, equating over three years to 771 hours of consultant neuroradiologist time or ~£40,000 (>$50,000) in NHS finance terms, given the workload of a neuro-oncology centre such as ours. Scaled across the UK, the saving is anticipated to be more than £1.5 million ($1.9 million). Framed differently, such software would enable these workforce hours to be reallocated to other areas of unmet clinical need. We furthermore show that patient survival forecasting is non-inferior when using these automated models, demonstrating the preservation of feature fidelity.

### Adding value with AI-assisted practice

Despite being well-validated in research to provide well-structured information on the imaging appearances of glioma, presenting an opportunity for quantitative tumour surveillance, the VASARI feature set is seldom used in clinical practice. The causes for this are multifactorial but are likely a combination of high clinical workload—VASARI is time-consuming to record—and lack of sufficient level of neuroradiology training and experience. Our software substantially lowers the barriers to adopting VASARI scoring while maintaining fidelity and assuring patient equity. Its introduction provides a means of extracting more detailed patient-personalized information, aiming to improve clinical care at a very modest cost in either time or financial terms. Particularly pertinent in the UK, where the number of radiologists per 100,000 population is one of the lowest in Western Europe(Piorkowska et al., 2017)—only 7—such decision support tools add high value or even free up an already overstretched workforce to allow work in other clinical areas.

A critical measure of the value of any feature, automated or manual, is downstream utility, such as survival prediction. Our analysis shows non-inferior—rather, quantitatively higher—predictive fidelity in using VASARI-auto features over those curated by consultant neuroradiologists. Demonstrating non-inferiority in software that is resource-cheap, contrasted with a time-consuming process for experienced neuroradiologists, is essential for software that provides inferior care than the current clinical standard, regardless of any efficiency or cost saving, adds little value.

## Maximising reporting consistency

Clinicians' opinions—whether radiologists or others—often differ. This is to be expected: diseases are typically heterogeneous. Patients, too, are heterogeneous: a successful treatment approach for one might not be suitable for another(Rajpurkar et al., 2022). However, a model capable of absorbing heterogeneity can yield a quantitative description that exhibits consistency across the population concerning a critical decision. From follow-up monitoring of tumours, we know that conventional two-dimensional measurements can be highly inconsistent between radiologists(Dempsey et al., 2005; McNitt-Gray et al., 2015; Zhao et al., 2009), motivating the pursuit of alternative approaches. Though we find it unlikely that a radiologist's work will be replaced entirely by software, harmonising human domain expertise with software-driven quantitative analytics seems inevitable to advance the clinical status quo. Notably, many comparisons could be drawn between this viewpoint and the commercial sector, where substantially greater AI development is currently undertaken. While car manufacturers increasingly navigate towards AI-assisted driving, steering wheels are unlikely to be removed anytime soon.

What are the characteristics of an optimal approach? The ideal would be to absorb all variation irrelevant to the task. For example, where measurements are undertaken by manual annotation—which one should note is the currently adopted clinical practice globally, despite their empirically observed limitations(Dempsey et al., 2005; McNitt-Gray et al., 2015; Zhao et al., 2009)—this is trivial to ameliorate using automated software and relatively simple mathematics. We exemplify this here, showing only modest inter-rater agreement between highly experienced neuroradiologists that can be stabilised and standardised with automated methods. Particularly pertinent examples are in deriving VASARI features (or, for that matter, any other radiological feature outside the scope of this article) that are ultimately quantitative. Where the quality of a lesion segmentation is validated, such as we show in our comparison between source segmentations and those in the segmentation model, then the mathematical derivation of precise proportions of lesional compartments, such as enhancing tumour, nonenhancing tumour, and perilesional signal change, is a simple mathematical operation of compartmental ratios. This is especially true for quantitative features that are harder to quantify intuitively, such as the thickness of enhancing tumour. Gliomas—particularly glioblastoma—are highly variable in their appearance. One part of an enhancing margin (if any) might be considerably thicker than another: how do we measure this? We would argue that the wrong answer (although commonplace in clinical reporting) would be to hedge an approximation between the lower and upper limits. Instead, a more robust solution is a simple mathematical derivation operating on a lesion segmentation(Ruffle et al., 2023a). However, the difficulty one faces, as is evidenced in these works, is where ambiguity in the ground truth—namely, what is a nonenhancing tumour and what is oedema—compounds an assessment regardless of whether derived by an experienced neuroradiologist or by software. An answer to this problem is unlikely to be solved by clinical experience, status quo imaging techniques or software, but rather by innovation across all three.

## Maximising performance equity

Healthcare should be equitable, which extends to any such tool at our clinical disposal(Abramoff et al., 2023; Carruthers et al., 2022). Artificial intelligence is one of the domains seeing the quickest growth in all research, industry, and society, with many purported applications across medicine. Yet equitable calibration to ensure that software brings benefits to all is relatively rarely quantified. For these reasons, we assess performance equity, not only of VASARI-auto but also of the adopted tumour segmentation model. Though confined to age and sex, the approach can be scaled through representation learning to encompass any characteristic(Carruthers et al., 2022).

## Limitations

Our study has limitations. First, although drawn from a larger cohort of 1172, we utilise a sample of 100 patients with glioma who have undergone comprehensive clinical VASARI featurisation by experienced neuroradiologists. Although large for the domain, further validation should be undertaken at a greater scale to evaluate broader generalisability. This sample, however, is carefully curated and includes imaging from two major US medical centres, for which we could evaluate the performance of the tumour segmentation model and VASARI-auto, both separately and taken together.

Second, we could not incorporate all features of VASARI in the software, specifically those that required structural neuroimaging, additional sequences beyond that provided by external repositories, or where variability/confabulation could occur. This decision was deliberate, for we wished to develop software that did not use patient-identifiable data, heterogeneous sequence data (in time and place), or computationally intensive processing pipelines. Whilst this choice precludes assessment of some of the VASARI features, the current pipeline requires merely a lesion segmentation to our strength. Moreover, VASARI-auto can be undertaken in a privacy-preserving setting, with trivial computing requirements, and from data acquired in any MRI. The appeal here is that rather than be limited to specific MRI scanners or specialist centres, the framework is immediately scalable to any centre, even with limited hardware resources. Future work should, however, expand upon this to include these remaining features.

Thirdly, the extent of the VASARI-auto featurisation pipeline was gated by the availability of widely used lesion compartment labels, namely enhancing tumour, nonenhancing tumour, and oedema(Baid et al., 2021). Therefore, our software could not provide VASARI data on haemorrhagic change because no label exists in the source data(Bakas et al., 2022b; Calabrese et al., 2022): it cannot learn what it has not been taught(Ruffle et al., 2023a). There are evolving opinions across neuro-oncology as to what may be oedema and what is nonenhancing tumour: it is for this reason we use the terminology of 'perilesional signal change' in discussing the segmentation pipeline. Moreover, it is the reason for lower agreement between VASARI-auto and neuroradiologist reporting for the lesion proportion features, since the software is guided by the status quo where such perilesional signal is referred to as oedema, though some radiologists might instead label as nonenhancing tumour. These changing viewpoints are because what classically was referred to as oedema has been shown to contain tumour cells under biopsy(Barajas et al., 2012). This ambiguity will impact model performance, for we are gated by ground truth labels that discretise nonenhancing tumour and perilesional signal change based on structural MRI sequences, despite there being no gold standard test to confirm if tumoural cells are definitively present within the signal abnormality or not. Though no 'silver bullet' imaging technique currently exists to remedy this, advanced imaging techniques (including diffusion and perfusion) may aid in remedying this in the future(Alsulami et al., 2023; Soni et al., 2018; Wurtemberger et al., 2022). In any case, it should be stressed that the values themselves do not actually matter here, but of far greater importance is that from lesion segmentation, a more robust and standardised assessment across a cohort of patients is yielded, likely the reason for stronger performance in downstream survival prediction.

Fourthly, since our accuracy metric is quantified from the neuroradiologist label as its ground truth, it would not be appropriate to claim superior accuracy of the VASARI-auto over a neuroradiologist here. However, we can quantify consistency between neuroradiologists, between the software and the radiologist, and between different input types of the software, which is akin to quantifying uncertainty. To that end, we illustrate that deriving VASARI features such as the thickness of the enhancing tumour margin and the presence of satellite lesions were radically inconsistent between neuroradiologists; these were far more reproducible between VASARI-auto runs.

Finally, our economic cost analysis assumes a VASARI feature set is undertaken in all cases assigned to the neuro-oncology MDT. This is an upper bound: VASARI is seldom used for the time and level of professional training it demands. Furthermore, given the sharp rise in the volume of medical imaging undertaken for patients globally(Piorkowska et al., 2017; Smith-Bindman et al., 2008), it is likely that the hours and cost incurred for radiologists to featurise these cases are a significant under-representation. However, precisely to that point, one should consider the economic analysis to highlight a gain in healthcare value at negligible time or financial cost.

## Conclusions

VASARI-auto can characterise glioma efficiently, effectively, and equitably. The use of VASARI-auto-derived features in predicting patient survival is non-inferior to the use of those manually curated by experienced consultant neuroradiologists. Translation to the clinical frontline with an automated derivation of these features may enhance existing radiology practice with negligible cost to an imaging department, serving as a decision support tool to provide healthcare providers with more information to facilitate standardised, equitable, and more personalised patient care.

# References

Abramoff, M.D., Tarver, M.E., Loyo-Berrios, N., Trujillo, S., Char, D., Obermeyer, Z., Eydelman, M.B., Foundational Principles of Ophthalmic, I., Algorithmic Interpretation Working Group of the Collaborative Community for Ophthalmic Imaging Foundation, W.D.C., Maisel, W.H., 2023. Considerations for addressing bias in artificial intelligence for health equity. NPJ Digit Med 6, 170. 10.1038/s41746-023-00913-9

Alsulami, T.A., Hyare, H., Thomas, D.L., Golay, X., 2023. The value of arterial spin labelling (ASL) perfusion MRI in the assessment of post-treatment progression in adult glioma: A systematic review and meta-analysis. Neurooncol Adv 5, vdad122. 10.1093/noajnl/vdad122

Antonelli, M., Reinke, A., Bakas, S., Farahani, K., Kopp-Schneider, A., Landman, B.A., Litjens, G., Menze, B., Ronneberger, O., Summers, R.M., van Ginneken, B., Bilello, M., Bilic, P., Christ, P.F., Do, R.K.G., Gollub, M.J., Heckers, S.H., Huisman, H., Jarnagin, W.R., McHugo, M.K., Napel, S., Pernicka, J.S.G., Rhode, K., Tobon-Gomez, C., Vorontsov, E., Meakin, J.A., Ourselin, S., Wiesenfarth, M., Arbelaez, P., Bae, B., Chen, S., Daza, L., Feng, J., He, B., Isensee, F., Ji, Y., Jia, F., Kim, I., Maier-Hein, K., Merhof, D., Pai, A., Park, B., Perslev, M., Rezaiifar, R., Rippel, O., Sarasua, I., Shen, W., Son, J., Wachinger, C., Wang, L., Wang, Y., Xia, Y., Xu, D., Xu, Z., Zheng, Y., Simpson, A.L., Maier-Hein, L., Cardoso, M.J., 2022. The Medical Segmentation Decathlon. Nat Commun 13, 4128. 10.1038/s41467-022-30695-9

Ashburner, J., Friston, K.J., 1999. Nonlinear spatial normalization using basis functions. Hum Brain Mapp 7, 254-266. 10.1002/(SICI)1097-0193(1999)7:4<254::AID-HBM4>3.0.CO;2-G

Ashburner, J., Friston, K.J., 2005. Unified segmentation. NeuroImage 26, 839-851. 10.1016/j.neuroimage.2005.02.018

Baid, U., Ghodasara, S., Bilello, M., Mohan, S., Calabrese, E., Colak, E., Farahani, K., Kalpathy-Cramer, J., Kitamura, F., Pati, S., Prevedello, L., Rudie, J., Sako, C., Shinohara, R., Bergquist, T., Chai, R., Eddy, J., Elliott, J., Reade, W., Bakas, S., 2021. The RSNA-ASNR-MICCAI BraTS 2021 Benchmark on Brain Tumor Segmentation and Radiogenomic Classification.

Bakas, S., Sako, C., Akbari, H., Bilello, M., Sotiras, A., Shukla, G., Rudie, J.D., Santamaría, N.F., Kazerooni, A.F., Pati, S., Rathore, S., Mamourian, E., Ha, S.M., Parker, W., Doshi, J., Baid, U., Bergman, M., Binder, Z.A., Verma, R., Lustig, R.A., Desai, A.S., Bagley, S.J., Mourelatos, Z., Morrissette, J., Watt, C.D., Brem, S., Wolf, R.L., Melhem, E.R., Nasrallah, M.P., Mohan, S., O'Rourke, D.M., Davatzikos, C., 2022a. Multi-parametric magnetic resonance imaging (mpMRI) scans for de novo Glioblastoma (GBM) patients from the University of Pennsylvania Health System (UPENN-GBM). The Cancer Imaging Archive. https://doi.org/10.7937/TCIA.709X-DN49

Bakas, S., Sako, C., Akbari, H., Bilello, M., Sotiras, A., Shukla, G., Rudie, J.D., Santamaría, N.F., Kazerooni, A.F., Pati, S., Rathore, S., Mamourian, E., Ha, S.M., Parker, W., Doshi, J., Baid, U., Bergman, M., Binder, Z.A., Verma, R., Lustig, R.A., Desai, A.S., Bagley, S.J., Mourelatos, Z., Morrissette, J., Watt, C.D., Brem, S., Wolf, R.L., Melhem, E.R., Nasrallah, M.P., Mohan, S.,

O'Rourke, D.M., Davatzikos, C., 2022b. The University of Pennsylvania glioblastoma (UPenn-GBM) cohort: advanced MRI, clinical, genomics, & radiomics. Scientific Data 9, 453. 10.1038/s41597-022-01560-7

Barajas, R.F., Jr., Phillips, J.J., Parvataneni, R., Molinaro, A., Essock-Burns, E., Bourne, G., Parsa, A.T., Aghi, M.K., McDermott, M.W., Berger, M.S., Cha, S., Chang, S.M., Nelson, S.J., 2012. Regional variation in histopathologic features of tumor specimens from treatment-naive glioblastoma correlates with anatomic and physiologic MR Imaging. Neuro Oncol 14, 942-954. 10.1093/neuonc/nos128

Biswas, A., Amirabadi, A., Wagner, M.W., Ertl-Wagner, B.B., 2022. Features of Visually AcceSAble Rembrandt Images: Interrater Reliability in Pediatric Brain Tumors. AJNR Am J Neuroradiol 43, 304-308. 10.3174/ajnr.A7399

Brett, Matthew, Markiewicz, Hanke, C., 2020. nipy/nibabel: 3.2.1 (Version 3.2.1). Zenodo. http://doi.org/10.5281/zenodo.4295521

Calabrese, E., Villanueva-Meyer, J.E., Rudie, J.D., Rauschecker, A.M., Baid, U., Bakas, S., Cha, S., Mongan, J.T., Hess, C.P., 2022. The University of California San Francisco Preoperative Diffuse Glioma MRI Dataset. Radiol Artif Intell 4, e220058. 10.1148/ryai.220058

Carruthers, R., Straw, I., Ruffle, J.K., Herron, D., Nelson, A., Bzdok, D., Fernandez-Reyes, D., Rees, G., Nachev, P., 2022. Representational ethical model calibration. NPJ Digit Med 5, 170. 10.1038/s41746-022-00716-4

Collins, G.S., Dhiman, P., Andaur Navarro, C.L., Ma, J., Hooft, L., Reitsma, J.B., Logullo, P., Beam, A.L., Peng, L., Van Calster, B., van Smeden, M., Riley, R.D., Moons, K.G., 2021. Protocol for development of a reporting guideline (TRIPOD-AI) and risk of bias tool (PROBAST-AI) for diagnostic and prognostic prediction model studies based on artificial intelligence. BMJ Open 11, e048008. 10.1136/bmjopen-2020-048008

Deeley, M.A., Chen, A., Datteri, R., Noble, J.H., Cmelak, A.J., Donnelly, E.F., Malcolm, A.W., Moretti, L., Jaboin, J., Niermann, K., Yang, E.S., Yu, D.S., Yei, F., Koyama, T., Ding, G.X., Dawant, B.M., 2011. Comparison of manual and automatic segmentation methods for brain structures in the presence of space-occupying lesions: a multi-expert study. Phys Med Biol 56, 4557-4577. 10.1088/0031-9155/56/14/021

Dempsey, M.F., Condon, B.R., Hadley, D.M., 2005. Measurement of tumor "size" in recurrent malignant glioma: 1D, 2D, or 3D? AJNR Am J Neuroradiol 26, 770-776. https://www.ncbi.nlm.nih.gov/pubmed/15814919

Gemini, L., Tortora, M., Giordano, P., Prudente, M.E., Villa, A., Vargas, O., Giugliano, M.F., Somma, F., Marchello, G., Chiaramonte, C., Gaetano, M., Frio, F., Di Giorgio, E., D'Avino, A., Tortora, F., D'Agostino, V., Negro, A., 2023. Vasari Scoring System in Discerning between Different Degrees of Glioma and IDH Status Prediction: A Possible Machine Learning Application? J Imaging 9. 10.3390/jimaging9040075

Gusev, Y., Bhuvaneshwar, K., Song, L., Zenklusen, J.C., Fine, H., Madhavan, S., 2018. The REMBRANDT study, a large collection of genomic data from brain cancer patients. Sci Data 5, 180158. 10.1038/sdata.2018.158

Harris, C.R., Millman, K.J., van der Walt, S.J., Gommers, R., Virtanen, P., Cournapeau, D., Wieser, E., Taylor, J., Berg, S., Smith, N.J., Kern, R., Picus, M., Hoyer, S., van Kerkwijk, M.H., Brett, M., Haldane, A., del Río, J.F., Wiebe, M., Peterson, P., Gérard-Marchant, P., Sheppard, K., Reddy, T., Weckesser, W., Abbasi, H., Gohlke, C., Oliphant, T.E., 2020. Array programming with NumPy. Nature 585, 357-362. 10.1038/s41586-020-2649-2

Hunter, J.D., 2007. Matplotlib: A 2D Graphics Environment. Computing in Science & Engineering 9, 90-95. 10.1109/MCSE.2007.55

Isensee, F., Jaeger, P.F., Full, P.M., Vollmuth, P., Maier-Hein, K., 2020. nnU-Net for Brain Tumor Segmentation. BrainLes@MICCAI.

Isensee, F., Jaeger, P.F., Kohl, S.A.A., Petersen, J., Maier-Hein, K.H., 2021. nnU-Net: a self-configuring method for deep learning-based biomedical image segmentation. Nat Methods 18, 203-211. 10.1038/s41592-020-01008-z

Jain, R., Poisson, L.M., Gutman, D., Scarpace, L., Hwang, S.N., Holder, C.A., Wintermark, M., Rao, A., Colen, R.R., Kirby, J., Freymann, J., Jaffe, C.C., Mikkelsen, T., Flanders, A., 2014. Outcome prediction in patients with glioblastoma by using imaging, clinical, and genomic biomarkers: focus on the nonenhancing component of the tumor. Radiology 272, 484-493. 10.1148/radiol.14131691

Lenchik, L., Heacock, L., Weaver, A.A., Boutin, R.D., Cook, T.S., Itri, J., Filippi, C.G., Gullapalli, R.P., Lee, J., Zagurovskaya, M., Retson, T., Godwin, K., Nicholson, J., Narayana, P.A., 2019. Automated Segmentation of Tissues Using CT and MRI: A Systematic Review. Acad Radiol 26, 1695-1706. 10.1016/j.acra.2019.07.006

Li, L., Fu, Y., Zhang, Y., Mao, Y., Huang, D., Yi, X., Wang, J., Tan, Z., Jiang, M., Chen, B.T., 2023. Magnetic resonance imaging findings of intracranial extraventricular ependymoma: A retrospective multi-center cohort study of 114 cases. Cancer Med 12, 16195-16206. 10.1002/cam4.6279

Louis, D.N., Perry, A., Wesseling, P., Brat, D.J., Cree, I.A., Figarella-Branger, D., Hawkins, C., Ng, H.K., Pfister, S.M., Reifenberger, G., Soffietti, R., von Deimling, A., Ellison, D.W., 2021. The 2021 WHO Classification of Tumors of the Central Nervous System: a summary. Neuro Oncol 23, 1231-1251. 10.1093/neuonc/noab106

Lu, S.-L., Xiao, F.-R., Cheng, J.C.-H., Yang, W.-C., Cheng, Y.-H., Chang, Y.-C., Lin, J.-Y., Liang, C.-H., Lu, J.-T., Chen, Y.-F., Hsu, F.-M., 2021. Randomized multi-reader evaluation of automated detection and segmentation of brain tumors in stereotactic radiosurgery with deep neural networks. Neuro-Oncology 23, 1560-1568. 10.1093/neuonc/noab071

Mandal, A.S., Romero-Garcia, R., Hart, M.G., Suckling, J., 2020. Genetic, cellular, and connectomic characterization of the brain regions commonly plagued by glioma. Brain 143, 3294-3307. 10.1093/brain/awaa277

McNitt-Gray, M.F., Kim, G.H., Zhao, B., Schwartz, L.H., Clunie, D., Cohen, K., Petrick, N., Fenimore, C., Lu, Z.Q., Buckler, A.J., 2015. Determining the Variability of Lesion Size Measurements from CT Patient Data Sets Acquired under "No Change" Conditions. Transl Oncol 8, 55-64. 10.1016/j.tranon.2015.01.001

Menze, B.H., Jakab, A., Bauer, S., Kalpathy-Cramer, J., Farahani, K., Kirby, J., Burren, Y., Porz, N., Slotboom, J., Wiest, R., Lanczi, L., Gerstner, E., Weber, M.A., Arbel, T., Avants, B.B., Ayache, N., Buendia, P., Collins, D.L., Cordier, N., Corso, J.J., Criminisi, A., Das, T., Delingette, H., Demiralp, C., Durst, C.R., Dojat, M., Doyle, S., Festa, J., Forbes, F., Geremia, E., Glocker, B., Golland, P., Guo, X., Hamamci, A., Iftekharuddin, K.M., Jena, R., John, N.M., Konukoglu, E., Lashkari, D., Mariz, J.A., Meier, R., Pereira, S., Precup, D., Price, S.J., Raviv, T.R., Reza, S.M., Ryan, M., Sarikaya, D., Schwartz, L., Shin, H.C., Shotton, J., Silva, C.A., Sousa, N., Subbanna, N.K., Szekely, G., Taylor, T.J., Thomas, O.M., Tustison, N.J., Unal, G., Vasseur, F., Wintermark, M., Ye, D.H., Zhao, L., Zhao, B., Zikic, D., Prastawa, M., Reyes, M., Van Leemput, K., 2015. The Multimodal Brain Tumor Image Segmentation Benchmark (BRATS). IEEE Trans Med Imaging 34, 1993-2024. 10.1109/TMI.2014.2377694

Nachev, P., Coulthard, E., Jager, H.R., Kennard, C., Husain, M., 2008. Enantiomorphic normalization of focally lesioned brains. NeuroImage 39, 1215-1226. 10.1016/j.neuroimage.2007.10.002

NHS, 2024. Guide to NHS waiting times in England. https://www.nhs.uk/nhs-services/hospitals/guide-to-nhs-waiting-times-in-england/

NHS-Employers, 2023. Pay and Conditions Circular (M&D) 4/2023. https://www.nhsemployers.org/system/files/2023-08/Pay%20and%20Conditions%20Circular%20%28MD%29%204-2023%20FINAL_0.pdf

Nicolasjilwan, M., Hu, Y., Yan, C., Meerzaman, D., Holder, C.A., Gutman, D., Jain, R., Colen, R., Rubin, D.L., Zinn, P.O., Hwang, S.N., Raghavan, P., Hammoud, D.A., Scarpace, L.M., Mikkelsen, T., Chen, J., Gevaert, O., Buetow, K., Freymann, J., Kirby, J., Flanders, A.E., Wintermark, M., Group, T.G.P.R., 2015. Addition of MR imaging features and genetic biomarkers strengthens glioblastoma survival prediction in TCGA patients. J Neuroradiol 42, 212-221. 10.1016/j.neurad.2014.02.006

Nilearn-contributors, 2024. nilearn. https://doi.org/10.5281/zenodo.8397156

Park, C.J., Han, K., Kim, H., Ahn, S.S., Choi, D., Park, Y.W., Chang, J.H., Kim, S.H., Cha, S., Lee, S.K., 2021. MRI Features May Predict Molecular Features of Glioblastoma in Isocitrate Dehydrogenase Wild-Type Lower-Grade Gliomas. AJNR Am J Neuroradiol 42, 448-456. 10.3174/ajnr.A6983

Paszke, A., Gross, S., Massa, F., Lerer, A., Bradbury, J., Chanan, G., Killeen, T., Lin, Z., Gimelshein, N., Antiga, L., Demsmaison, A., Köpf, A., Yang, E., DeVito, Z., Raison, M., Tejani, A., Chilamkurthy, S., Steiner, B., Fang, L., Bai, J., Chintala, S., 2019. PyTorch: An Imperative Style, High-Performance Deep Learning Library. NeurIPS.

Pedregosa, F., Varoquaux, G., Gramfort, A., 2011. Scikit-learn: Machine Learning in Python. Journal of Machine Learning Research 12, 2825-2830.

Peeken, J.C., Goldberg, T., Pyka, T., Bernhofer, M., Wiestler, B., Kessel, K.A., Tafti, P.D., Nusslin, F., Braun, A.E., Zimmer, C., Rost, B., Combs, S.E., 2019. Combining multimodal imaging and treatment features improves machine learning-based prognostic assessment in patients with glioblastoma multiforme. Cancer Med 8, 128-136. 10.1002/cam4.1908

Peeken, J.C., Hesse, J., Haller, B., Kessel, K.A., Nusslin, F., Combs, S.E., 2018. Semantic imaging features predict disease progression and survival in glioblastoma multiforme patients. Strahlenther Onkol 194, 580-590. 10.1007/s00066-018-1276-4

Peng, J., Kim, D.D., Patel, J.B., Zeng, X., Huang, J., Chang, K., Xun, X., Zhang, C., Sollee, J., Wu, J., Dalal, D.J., Feng, X., Zhou, H., Zhu, C., Zou, B., Jin, K., Wen, P.Y., Boxerman, J.L., Warren, K.E., Poussaint, T.Y., States, L.J., Kalpathy-Cramer, J., Yang, L., Huang, R.Y., Bai, H.X., 2021. Corrigendum to: Deep learning-based automatic tumor burden assessment of pediatric high-grade gliomas, medulloblastomas, and other leptomeningeal seeding tumors. Neuro-Oncology 23, 2124-2124. 10.1093/neuonc/noab226

Piorkowska, M., Goh, V., Booth, T.C., 2017. Post Brexit: challenges and opportunities for radiology beyond the European Union. Br J Radiol 90, 20160852. 10.1259/bjr.20160852

Rajpurkar, P., Chen, E., Banerjee, O., Topol, E.J., 2022. AI in health and medicine. Nature Medicine. 10.1038/s41591-021-01614-0

Reback, J., McKinney, W., jbrockmendel, 2020. pandas-dev/pandas: Pandas 1.0.3 (Version v1.0.3). Zenodo. http://doi.org/10.5281/zenodo.3715232

Ritaccio, A.L., Brunner, P., Schalk, G., 2018. Electrical Stimulation Mapping of the Brain: Basic Principles and Emerging Alternatives. J Clin Neurophysiol 35, 86-97. 10.1097/WNP.0000000000000440

Ruffle, J., K., Mohinta, S., Gray, R., Hyare, H., Nachev, P., 2023a. Brain tumour segmentation with incomplete imaging data. Brain Commun. 10.1093/braincomms/fcad118

Ruffle, J.K., 2023. Brain tumour segmentation with incomplete imaging data. GitHub. https://github.com/high-dimensional/tumour-seg

Ruffle, J.K., Mohinta, S., Pombo, G., Gray, R., Kopanitsa, V., Lee, F., Brandner, S., Hyare, H., Nachev, P., 2023b. Brain tumour genetic network signatures of survival. Brain. https://ui.adsabs.harvard.edu/abs/2023arXiv230106111R

Setyawan, N.H., Choridah, L., Nugroho, H.A., Malueka, R.G., Dwianingsih, E.K., 2024. Beyond invasive biopsies: using VASARI MRI features to predict grade and molecular parameters in gliomas. Cancer Imaging 24, 3. 10.1186/s40644-023-00638-8

Smith-Bindman, R., Miglioretti, D.L., Larson, E.B., 2008. Rising use of diagnostic medical imaging in a large integrated health system. Health Aff (Millwood) 27, 1491-1502. 10.1377/hlthaff.27.6.1491

Soni, N., Srindharan, K., Kumar, S., Mishra, P., Bathla, G., Kalita, J., Behari, S., 2018. Arterial spin labeling perfusion: Prospective MR imaging in differentiating neoplastic from non-neoplastic intra-axial brain lesions. Neuroradiol J 31, 544-553. 10.1177/1971400918783058

Suetens, P., Bellon, E., Vandermeulen, D., Smet, M., Marchal, G., Nuyts, J., Mortelmans, L., 1993. Image segmentation: methods and applications in diagnostic radiology and nuclear medicine. Eur J Radiol 17, 14-21. 10.1016/0720-048x(93)90023-g

sust-it.net, 2024. Energy Cost Calculator - UK: Price Cap (Jan 2024). https://www.sust-it.net/energy-calculator.php

TCIA, 2020. VASARI Research Project. https://wiki.cancerimagingarchive.net/display/Public/VASARI+Research+Project

The-MONAI-Consortium, 2020. Project MONAI. Zenodo. https://doi.org/10.5281/zenodo.4323059

Topol, E., 2019. The Topol Review: Preparing the healthcare workforce to deliver the digital future. In: NHS (Ed.). https://topol.hee.nhs.uk/wp-content/uploads/HEE-Topol-Review-2019.pdf

Wan, Y., Rahmat, R., Price, S.J., 2020. Deep learning for glioblastoma segmentation using preoperative magnetic resonance imaging identifies volumetric features associated with survival. Acta Neurochir (Wien) 162, 3067-3080. 10.1007/s00701-020-04483-7

Wang, J., Yi, X., Fu, Y., Pang, P., Deng, H., Tang, H., Han, Z., Li, H., Nie, J., Gong, G., Hu, Z., Tan, Z., Chen, B.T., 2021. Preoperative Magnetic Resonance Imaging Radiomics for Predicting Early Recurrence of Glioblastoma. Front Oncol 11, 769188. 10.3389/fonc.2021.769188

Waskom, M., Seaborn-Development-Team, 2020. seaborn. Zenodo. https://doi.org/10.5281/zenodo.4645478

Wurtemberger, U., Rau, A., Reisert, M., Kellner, E., Diebold, M., Erny, D., Reinacher, P.C., Hosp, J.A., Hohenhaus, M., Urbach, H., Demerath, T., 2022. Differentiation of Perilesional Edema in Glioblastomas and Brain Metastases: Comparison of Diffusion Tensor Imaging, Neurite Orientation Dispersion and Density Imaging and Diffusion Microstructure Imaging. Cancers (Basel) 15. 10.3390/cancers15010129

Xue, J., Wang, B., Ming, Y., Liu, X., Jiang, Z., Wang, C., Liu, X., Chen, L., Qu, J., Xu, S., Tang, X., Mao, Y., Liu, Y., Li, D., 2020. Deep learning–based detection and segmentation-assisted management of brain metastases. Neuro-Oncology 22, 505-514. 10.1093/neuonc/noz234

Zhao, B., James, L.P., Moskowitz, C.S., Guo, P., Ginsberg, M.S., Lefkowitz, R.A., Qin, Y., Riely, G.J., Kris, M.G., Schwartz, L.H., 2009. Evaluating variability in tumor measurements from same-day repeat CT scans of patients with non-small cell lung cancer. Radiology 252, 263-272. 10.1148/radiol.2522081593

Zhou, H., Vallieres, M., Bai, H.X., Su, C., Tang, H., Oldridge, D., Zhang, Z., Xiao, B., Liao, W., Tao, Y., Zhou, J., Zhang, P., Yang, L., 2017. MRI features predict survival and molecular markers in diffuse lower-grade gliomas. Neuro Oncol 19, 862-870. 10.1093/neuonc/now256

# Tables

| Table 1: Study cohort data | | | |
|---|---|---|---|
| **Parameter** | **All patients** | **Glioblastoma, IDH-wt cohort** | **VASARI labelled by consultant neuroradiologists** |
| Number of patients | 1172 | 1045 | 100 |
| Age (years) ± SD | 60 ± 13.84 | 62 ± 12.24 | 61 ± 13.49 |
| Sex | 704 male, 468 female | 627 male, 418 female | 56 male, 44 female |
| Molecular diagnosis | Glioblastoma, IDH-wt (89.16%), Astrocytoma, IDH-mut (9.73%), Oligodendroglioma, IDH-mut and 1p/19q co-deleted (1.11%). | Glioblastoma, IDH-wt (100%) | Glioblastoma, IDH-wt (100%) |
| Overall survival (days) | 503 ± 453.90 | 441 ± 370.29 | 436 ± 462.21 |

| Feature | Coefficient | Std Error | t | P>\|t\| | [0.025 | 0.975] |
|---|---|---|---|---|---|---|
| Constant | 0.532 | 1.035 | 0.514 | 0.609 | -1.531 | 2.595 |
| F4 Enhancement Quality | 0.050 | 0.078 | 0.645 | 0.521 | -0.105 | 0.205 |
| F5 Proportion of Enhancing Tumour | -0.106 | 0.077 | -1.376 | 0.173 | -0.260 | 0.048 |
| F14 Proportion of Oedema | 0.072 | 0.189 | 0.379 | 0.706 | -0.305 | 0.448 |
| F1 Tumour Location (Temporal) | -0.078 | 0.083 | -0.938 | 0.351 | -0.243 | 0.087 |
| F1 Tumour Location (Insula) | -0.269 | 0.226 | -1.193 | 0.237 | -0.719 | 0.180 |
| F1 Tumour Location (Parietal) | -0.152 | 0.094 | -1.62 | 0.109 | -0.338 | 0.035 |
| F1 Tumour Location (Occipital) | -0.248 | 0.162 | -1.526 | 0.131 | -0.571 | 0.076 |
| F1 Tumour Location (Brainstem) | -0.207 | 0.327 | -0.632 | 0.529 | -0.860 | 0.445 |
| F1 Tumour Location (Corpus Callosum) | -0.117 | 0.155 | -0.750 | 0.456 | -0.426 | 0.193 |
| F9 Multifocal | 0.225 | 0.178 | 1.265 | 0.210 | -0.130 | 0.580 |
| F19 Ependymal Invasion | 0.028 | 0.098 | 0.282 | 0.779 | -0.168 | 0.224 |
| F21 Deep WM Invasion | -0.213 | 0.095 | -2.239 | 0.028 | -0.402 | -0.023 |

Supplementary Table 1. Coefficients to VASARI-auto survival model.